\documentclass[onecolumn]{aastex631}
\usepackage{natbib}
\usepackage{amsmath}
\usepackage{bm}
\usepackage{graphicx}
\usepackage{multirow}
\usepackage{booktabs} 
\usepackage{array}
\usepackage{xcolor} 
\usepackage{hyperref}
\usepackage{soul}

\begin{document}

\title{Non-thermal Sources from Stereoscopic Hard X-ray and Earth-based Microwave Observations in a Data-Constrained Magnetohydrodynamic Simulation}

\author[0000-0003-2002-0247]{Keitarou Matsumoto}
\affiliation{Center for Solar-Terrestrial Research, New Jersey Institute of Technology, Newark, NJ 07102-1982, USA}

\affiliation{Institute for Space-Earth Environmental Research, Nagoya University, Furo-cho, Chikusa-ku, Nagoya,
Aichi 464-8601, Japan}

\author[0000-0001-5121-5122]{Satoshi Inoue}
\affiliation{Center for Solar-Terrestrial Research, New Jersey Institute of Technology, Newark, NJ 07102-1982, USA}

\author[0000-0002-2633-3562
]{Meiqi Wang}
\affiliation{Center for Solar-Terrestrial Research, New Jersey Institute of Technology, Newark, NJ 07102-1982, USA}

\author[0000-0002-0660-3350]{Bin Chen}
\affiliation{Center for Solar-Terrestrial Research, New Jersey Institute of Technology, Newark, NJ 07102-1982, USA}

\author[0000-0002-8538-3455]{Muriel Zoë Stiefel}
\affiliation{University of Applied Sciences and Arts Northwestern Switzerland, Bahnhofstrasse 6, 5210 Windisch, Switzerland}
\affiliation{ETH Zürich, Rämistrasse 101, 8092 Zürich, Switzerland}

\author[0000-0002-2002-9180]{Säm Krucker}
\affiliation{University of Applied Sciences and Arts Northwestern Switzerland, Bahnhofstrasse 6, 5210 Windisch, Switzerland}
\affiliation{Space Sciences Laboratory, University of California, 7 Gauss Way, 94720 Berkeley, USA}

\author[0000-0001-5037-9758
]{Satoshi Masuda}
\affiliation{Institute for Space-Earth Environmental Research, Nagoya University, Furo-cho, Chikusa-ku, Nagoya,
Aichi 464-8601, Japan}

\author[0000-0002-5233-565X]{Haimin Wang}
\affiliation{Center for Solar-Terrestrial Research, New Jersey Institute of Technology, Newark, NJ 07102-1982, USA}

\email{km876@njit.edu}



\begin{abstract}
We analyze the X7.1 flare on 2024 October 1 from NOAA AR 13842 using hard X-ray (HXR) imaging, microwave observations by the Expanded Owens Valley Solar Array (EOVSA), and a three-dimensional Magnetohydrodynamic (MHD) simulation. The flare was observed from two vantage points, with Solar Orbiter/Spectrometer Telescope for Imaging X-rays viewing the flare near the limb and Advanced Space-based Solar Observatory/Hard X-ray Imager and EOVSA observing it on the disk. We carried out a data-constrained MHD simulation using a nonlinear force-free field extrapolation as the initial condition and constrained the height of the non-thermal looptop source from stereoscopic HXR and microwave observations. The height is consistent between the stereoscopic analysis and the MHD simulation. A secondary non-thermal microwave source aligned with a southward plasma ejection corresponds to an elongated current sheet. Although the current sheet grows in multiple directions, the secondary microwave emission is observed only from the southern segment. This localization suggests reconnection in regions with different magnetic field strengths. Reconnection in strong-field regions produces flare arcades with dominant looptop emission, whereas reconnection in weaker southern regions gives rise to secondary microwave emission at higher altitudes. The height of the secondary source is consistent between the stereoscopic analysis and the MHD simulation. Microwave spectral fitting suggests a higher low-energy cutoff for non-thermal electrons in the secondary microwave source than in the main looptop source. This may reflect the transport of electrons pre-accelerated near the looptop source by the southward plasma ejection.
\end{abstract}



\section{Introduction} \label{sec:intro}
In solar flares, accelerated electrons often produce hard X-rays (HXR) and microwave emissions, which are commonly used to diagnose non-thermal electrons, although both wavelength ranges can also include thermal components. The non-thermal HXR mainly originates from energetic electrons and is often produced at footpoints of coronal loops \citep{Fletcher2011}. The non-thermal HXR generated in the corona can be observed in space \citep{Masuda1994,Krucker2010,Chen2024}. In the GHz range, non-thermal microwave emission during flares is typically dominated by accelerated electrons trapped by magnetic field lines \citep{Bastian1998}. While the brightest microwave emission concentrates near the looptop region \citep{Sijie2020,Chen2024}, depending on frequency, they can appear in a wide range of coronal heights \citep{Chen2020}. In addition to the footpoints of the reconnected post-flare loops, recent studies have reported HXR and microwave sources located at the footpoints of erupting field lines, which are associated with the filament eruption \citep{Chen2020ApJ,Stiefel2023,Chen2025}. This scenario is also supported by a data-constrained magnetohydrodynamic (MHD) simulation \citep{Matsumoto2026}. However, the detailed processes of particle acceleration and transport associated with HXR and microwave emissions remain under debate.

To address particle acceleration and transport, it is essential to consider the three-dimensional (3D) configuration of the magnetic fields because the electrons are guided by the magnetic field lines. While we can observe only magnetic structures projected onto the sky plane, observations from two or more different viewing angles can be used to reconstruct the 3D geometry using a technique known as stereoscopy. Using this technique, two spacecraft of the Solar TErrestrial RElations Observatory (STEREO; \citealt{Kaiser2008}) reconstructed 3D coronal loops that are different from potential fields \citep{Aschwanden2008,Aschwanden2012}. Stereoscopic X-ray observations provide strong constraints on the 3D geometry of particle acceleration sites. For example, \citet{Ryan2024b} combined Solar Orbiter/STIX \citep{Krucker2020} with Hinode/XRT \citep{Golub2007}, extracting the 3D volume of the thermal X-ray source and indicating that the scaling law $V = A^{3/2}$, where $A$ and $V$ are the observed X-ray area and 3D derived volume, respectively, overestimates the true source volume. \citet{Ryan2024a} paired STIX with ASO-S/HXI \citep{Zhang2019}, reproducing the 3D coronal loops associated with the coronal HXR source.

Nonlinear force-free field (NLFFF) extrapolations are widely used to reconstruct 3D coronal magnetic fields \citep{Wiegelmann2012, Inoue2016}. In the low-$\beta$ corona ($\beta \sim 0.01$--0.1; \citealt{Gary2001}), where $\beta$ is the ratio of gas to magnetic pressure, the Lorentz force dominates, and the force-free assumption is well approximated. However, NLFFF provides only a static snapshot and cannot cover the eruptive evolution. Data-constrained MHD simulations overcome this limitation and have successfully reproduced eruptive dynamics and observations \citep{Yamasaki2022, Liu2025, Matsumoto2025a,Roddanavar2026}.

This paper is a continuation of the work by \citet{Matsumoto2026}, which showed that the data-constrained MHD simulation reproduces the eruption and the associated reconnected field lines linked to the observed HXR footpoint sources, illuminating a detailed reconnection process in a realistic 3D magnetic environment. In this paper, we combined stereoscopic HXR and microwave observations with a 3D data-constrained MHD simulation to constrain the height of the non-thermal coronal source and test the consistency with the MHD simulation. Section \ref{sec:sec2} describes the observations and numerical setup, Section \ref{sec:sec3} presents the main results, Section \ref{sec:sec4} provides the discussions, and Section \ref{sec:sec5} summarizes our conclusions.

\section{Observations and MHD simulations} \label{sec:sec2}
\subsection{Observation} \label{sec:sec2.1}
\begin{figure*}[!htbp]
\centering
\includegraphics[width=0.8\textwidth]{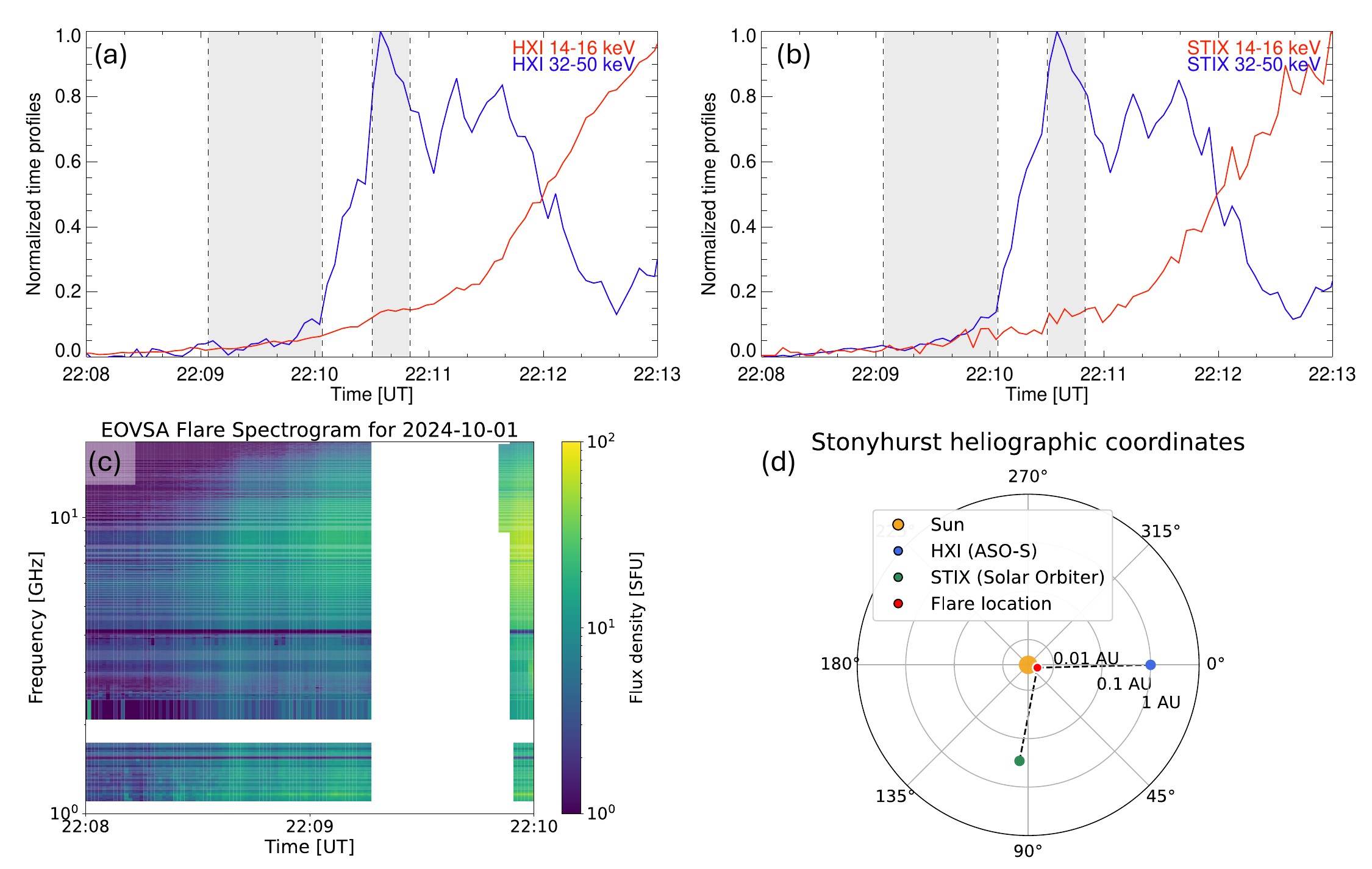}
\caption{(a) and (b) Temporal evolution of X-ray observed by HXI and STIX. For both HXI and STIX, the data are summed to a 4 s resolution. The red and blue curves represent the 14-16 keV and 32-50 keV ranges, respectively. The STIX light curves are shifted by 352.9 s to consider the different travel time of the light. The two gray shaded regions indicate the time intervals 22:09:04-22:10:04 (Phase 0) and 22:10:30-22:10:50 (Phase 1), which are used for hard X-ray imaging. (c) Microwave dynamic spectrum observed by EOVSA between 22:08 UT and 22:10 UT. White pixels show no data. (d) Stonyhurst heliographic coordinates show the locations of HXI, STIX, and the flare location. Separation angle between HXI and STIX is $\sim95^{\circ}$.
}\label{fig:fig1}
\end{figure*}

The X7.1 flare occurred in NOAA AR 13842 on 2024 October 1, starting at 21:58 UT and peaking at 22:20 UT in the GOES X-ray. \citet{Stiefel2025a} reported a super-hot component of the coronal source in the main peak phase of this flare (22:14:20–22:14:50 UT). Figures \ref{fig:fig1}(a) and (b) show the light curves from HXI and STIX. The STIX profile is built by summing eight pixels from the 24 coarsest imaging detectors, whereas the HXI profile is the sum of the two thick detectors (D92 and D93). Because the light travel times to HXI and STIX differ, the STIX time series is shifted by 352.9 s to align with HXI. The STIX attenuator moved at 22:04:42 UT (22:10:34 UT in Figure \ref{fig:fig1}). To obtain an unattenuated 14–16 keV curve for Figure \ref{fig:fig1}(b), we adopt the STIX background detector pixels 2 \& 5 \citep{Stiefel2025}. Figure \ref{fig:fig1}(c) shows the Expanded Owens Valley Solar Array (EOVSA; \citealt{Gary2018}) cross power spectrogram, with the white region indicating no data. Since 2017, EOVSA has provided microwave imaging spectroscopy spanning 1 to 18 GHz with 2 s time resolution, enabling spectral analysis at the flare site. After 22:10 UT, EOVSA data are unavailable due to phase calibration. We focus on the pre-impulsive and impulsive phases, defined by two intervals, 22:09:04-22:10:04 (Phase 0) and 22:10:30-22:10:50 (Phase 1), as marked in Figures \ref{fig:fig1}(a) and (b). Phase 0 with EOVSA observation is the main focus of this paper, whereas Phase 1 is discussed in detail in \citet{Matsumoto2026}. Figure \ref{fig:fig1}(d) shows the position of HXI, STIX, and the flare location. The HXI–STIX separation is $\sim95^{\circ}$, providing two complementary vantage points on the HXR emission.

A robust comparison of HXI and STIX images requires precise co-alignment. Following previous work (e.g., \citealt{Massa2022,Ryan2024a}), we register the HXR sources using ultraviolet (UV) ribbons observed by the Solar Dynamics Observatory (SDO)/ Atmospheric Imaging Assembly (AIA; \citealt{Lemen2012}). To align the HXI and STIX images, we applied the same shifts as used for Phase 1 in \citet{Matsumoto2026}. We used the same shift for the STIX image in Phase 0 because the main reconnection site responsible for the non-thermal emission is expected to remain unchanged between Phases 0 and 1, as discussed in Section \ref{sec:sec3.2}. HXI images are reconstructed with CLEAN, and STIX images with the maximum-entropy method MEM\_GE \citep{Massa2020}. As noted by \citealt{Massa2022}, MEM\_GE typically achieves a smaller $\chi^{2}$ relative to the observed visibilities than CLEAN, while both recover consistent large-scale morphologies. Since our aim is to locate the HXR footpoints rather than resolve fine structure, we restrict HXI to subcollimators $\ge 13.4^{\prime\prime}$ (D39–D91) and adopt a $20^{\prime\prime}$ beam for STIX. The resulting images are presented in Section \ref{sec:sec3}.

\subsection{Nonlinear Force-free Extrapolation}\label{sec:nlfff}
We performed an NLFFF extrapolation using SDO/ Helioseismic and Magnetic Imager (HMI; \citealt{Scherrer2012}) vector magnetogram at 20:36 UT (cylindrical equal-area projection) as the bottom boundary, following \citet{Matsumoto2025b}. The governing equations are
\begin{equation}
\mathit{\rho} = \lvert \bm{B} \rvert,
\label{eq:eq1}
\end{equation}
\begin{equation}
\frac{\partial \bm{v}}{\partial t} = -(\bm{v} \cdot \nabla)\bm{v} + \frac{1}{\mathit{\rho}} \bm{J} \times \bm{B} + \mathit{\nu} \nabla^2 \bm{v},
\label{eq:eq2}
\end{equation}
\begin{equation}
\frac{\partial \bm{B}}{\partial t} = \nabla \times (\bm{v} \times \bm{B})+\eta \nabla^2\bm{B} - \nabla \phi,
\label{eq:eq3}
\end{equation}
\begin{equation}
\bm{J} = \nabla \times \bm{B},
\label{eq:eq4}
\end{equation}
\begin{equation}
\frac{\partial \phi}{\partial t} + c_h^2 \nabla \cdot \bm{B} = - \frac{c_h^2}{c_p^2} \phi,
\label{eq:eq5}
\end{equation}
where $\rho$, $\bm{B}$, $\bm{v}$, $\bm{J}$, and $\phi$ denote plasma density, magnetic field, velocity, current density, and the scalar potential. Length, magnetic field, density, velocity, time, and current are normalized by $L^{*}=362.5$~Mm, $B^{*}=2.915\times10^{-1}$~T, $\rho^{*}$ (kg m$^{-3}$) at the bottom boundary, $V_A^{*}=B^{*}/(\mu_0 \rho^{*})^{1/2}$ (m s$^{-1}$), $\tau_A^{*}=L^{*}/V_A^{*}$ (s), and $J^{*}=B^{*}/(\mu_0 L^{*})$ (A m$^{-2}$), respectively. $\mu_0$ is the magnetic permeability. We adopt the pseudo density $\rho=\lvert\bm{B}\rvert$ to obtain a smoother spatial distribution of the Alfvén speed and facilitate the relaxation process \citep{Inoue2014a}. Divergence errors are controlled by the hyperbolic divergence cleaning \citep{Dedner2002} with $c_h^2=0.04$ and $c_p^2=0.1$. Viscosity and resistivity are $\nu=1.0\times10^{-3}$ and $\eta=5.0\times10^{-5}+1.0\times10^{-3}\,\lvert\bm{J}\times\bm{B}\rvert\,\lvert\bm{v}\rvert^{2}/\lvert\bm{B}\rvert^{2}$.

The initial field is the potential field from $B_z$ via the Green’s function method \citep{Sakurai1982}. Regarding the boundary conditions, the normal magnetic field is fixed on all faces, while the tangential components follow the induction equation except at the bottom. At the bottom, the tangential field is $\bm{B}_{\rm bc}=\gamma \bm{B}_{\rm obs} + (1-\gamma)\bm{B}_{\rm pot}$, where the parameter $\gamma\in[0,1]$, the observed magnetic field $\bm{B}_{\text{obs}}$, and the potential magnetic field $\bm{B}_{\text{pot}}$. We change $\gamma$ from 0 with step $d\gamma=0.02$ when $R=\int \lvert\bm{J}\times\bm{B}\rvert^{2} dV$ falls below $R_{\rm min}=5.0\times10^{-3}$. To suppress boundary artifacts, if the Alfv\'en Mach number $M_A=\lvert\bm{v}\rvert/\lvert\bm{v}_A\rvert$ exceeds $v_{\rm max}=0.04$, we rescale $\bm{v}\rightarrow (v_{\rm max}/M_A)\bm{v}$. Regarding the velocity and the potential at the boundaries,  velocities are zero, and $\partial\phi/\partial n=0$ is given at each boundary.

\subsection{Data-constrained MHD Simulation}\label{sec:mhd}
We then ran data-constrained MHD simulations starting from the NLFFF \citep{Inoue2016}. The equations are the same as above, but the boundary treatment and velocity limit differ. The normal magnetic field is fixed, while the tangential components evolve following the induction equation (Eq.(\ref{eq:eq3})) with zero velocity. Note that although we employ the same equation (Eq.(\ref{eq:eq1})) to describe the evolution of the plasma density, comparisons between simulations with and without solving the mass continuity equation show only minor quantitative differences, while preserving the topology and large-scale dynamics \citep{Inoue2014}. We set $\eta=1.0\times10^{-5}$ and $\nu=1.0\times10^{-4}$. The domain is $362.5^{3}$~Mm$^{3}$ ($1.0^{3}$ in normalized units). HMI data were binned from $1000^{2}$ to $500^{2}$. Time is normalized by the Alfv\'en time ($t=1.0$ corresponds to $\sim6$~min). For the MHD dynamics of this X7.1 flare revealed by \citet{Matsumoto2025b}, tether-cutting reconnection between sheared field lines forms a magnetic flux rope that erupts. The data-constrained MHD simulation reproduces the observational features of the X7.1 flare. We refer the interested readers to \citet{Matsumoto2025b}.

\section{Results} \label{sec:sec3}

\subsection{Reference Magnetic Field for Comparison with Observation}\label{sec:sec3.1}
To compare the observations with simulations, we construct synthetic flare ribbons from the simulation following \cite{Inoue2014}. By comparing with the AIA 1600 \AA, we identify the simulation time that best reproduces the observed ribbon morphology, and the corresponding magnetic field structure is used for the subsequent analysis.

Following \citet{Inoue2014}, we trace footpoints involved in magnetic reconnection using the change of the connectivity of the field lines. We denote the footpoint location by $\boldsymbol{x}_0$. Tracing the magnetic field line from $\boldsymbol{x}_0$ gives the conjugate footpoint
$\boldsymbol{x}_1(\boldsymbol{x}_0,t_n)$. The change in connectivity
between two successive times $t_{n-1}$ and $t_{n}$ is
\begin{equation}
  \delta(\boldsymbol{x}_0,t_n)
  \equiv \left\lvert
  \boldsymbol{x}_1(\boldsymbol{x}_0,t_{n})
  - \boldsymbol{x}_1(\boldsymbol{x}_0,t_{n-1})
  \right\rvert
\end{equation}
We evaluate $\delta$ only for lines with the magnetic twist number $T_w\ge 0.5$ to focus on the dynamics of the eruption \citep{Matsumoto2025b}.
The time-integrated measure for the displacement is,
\begin{equation}
  \Delta x(\boldsymbol{x}_0,t)
  \equiv \sum_{i=0}^{n} \delta(\boldsymbol{x}_0,t_i),
\end{equation}
records the history of connectivity changes at $\boldsymbol{x}_0$, indicating footpoints involved in the reconnection.

\begin{figure*}[!htbp]
\centering
\includegraphics[width=0.65\textwidth]{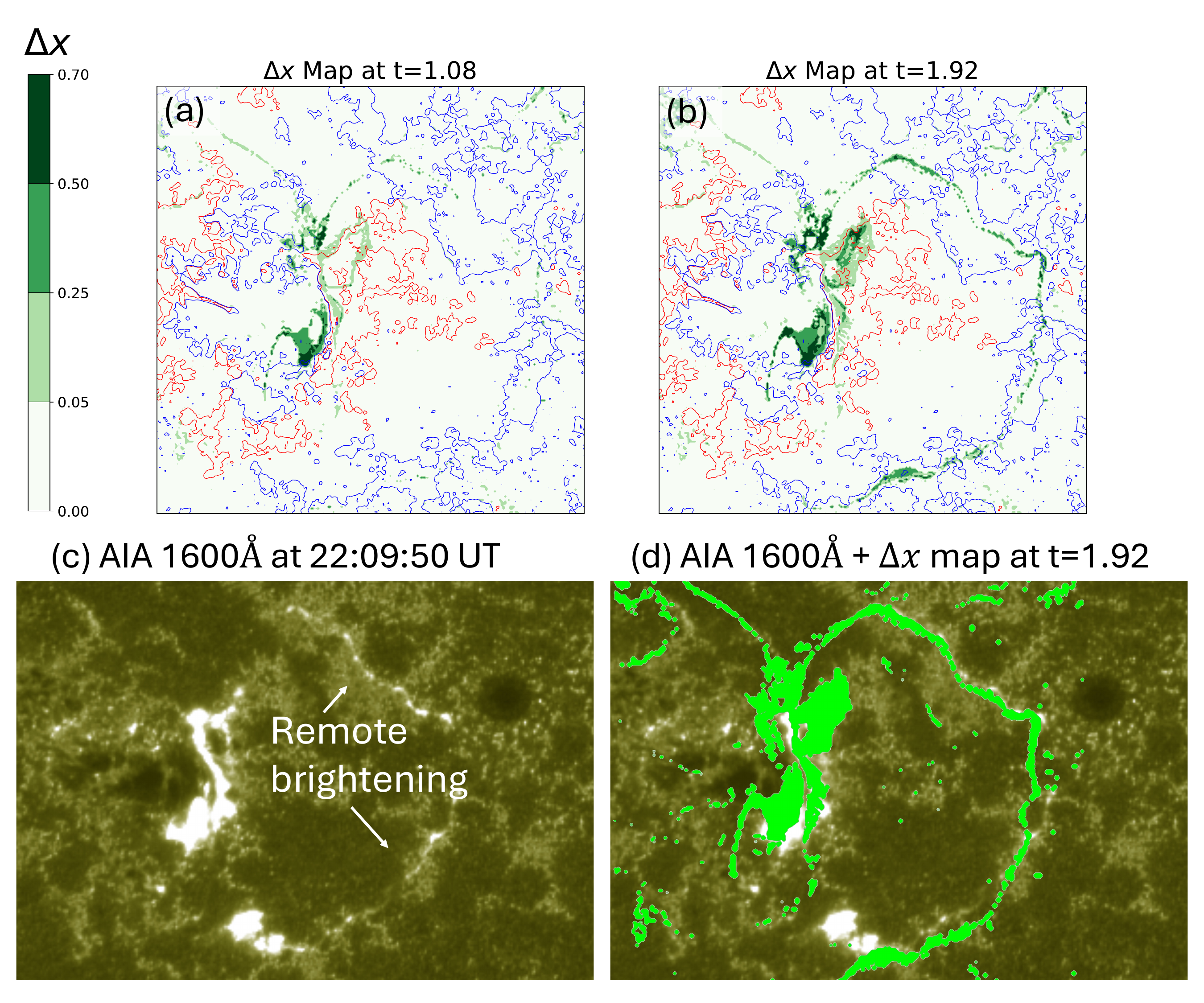}
\caption{(a) $\Delta x$ map at $t=1.08$. Red and blue contours mark $B_z=1.0\times10^{-2}$ T and $B_z=-5.0\times10^{-3}$ T at 20:36 UT. (b) $\Delta x$ map at $t=1.92$ with the same $B_z$ contours as in (a). (c)  AIA 1600 \AA\ at 22:09:50 UT highlighting the remote brightening. (d) Same as (c) with $\Delta x>0.05$ (green) at $t=1.92$ overlaid. The animation of panels (a) and (b) proceeds from $t = 0.00$ to $2.40$.
}\label{fig:fig2}
\end{figure*}

Figures \ref{fig:fig2}(a) and (b) show the $\Delta x$ maps at $t=1.08$ and $t=1.92$, highlighting the change of field line connectivity. At $t=1.08$, the reconnection site was concentrated along the polarity inversion line, while the remote region appeared in the west at $t=1.92$. Figures \ref{fig:fig2}(c) and (d) compare AIA 1600 \AA\ in Phase 0 with the $\Delta x$ map at $t=1.92$, evidencing correspondence at both flare ribbons and the remote brightening. Because the morphology of $\Delta x$ at $t \ge 1.92$ matches the remote brightening in Phase 0, we take the simulation state at and after $t=1.92$ as the appropriate reference for comparison with the observations in Phase 0. These results indicate that simulations with $t \ge 1.92$ are appropriate for comparison with the observations. The AIA morphology in Figure \ref{fig:fig2}(c), including the remote brightening, is already present by 22:08:30 UT. Therefore, we use $t=1.92$ for the 22:08:30 UT comparison (Sections \ref{sec:sec3.3} and \ref{sec:sec4.2}) and $t=2.4$ for the Phase 0 comparison (Section \ref{sec:sec3.4}).

\subsection{Hard X-ray and Microwave Images} \label{sec:sec3.2}

\begin{figure*}[!htbp]
\centering
\includegraphics[width=0.65\textwidth]{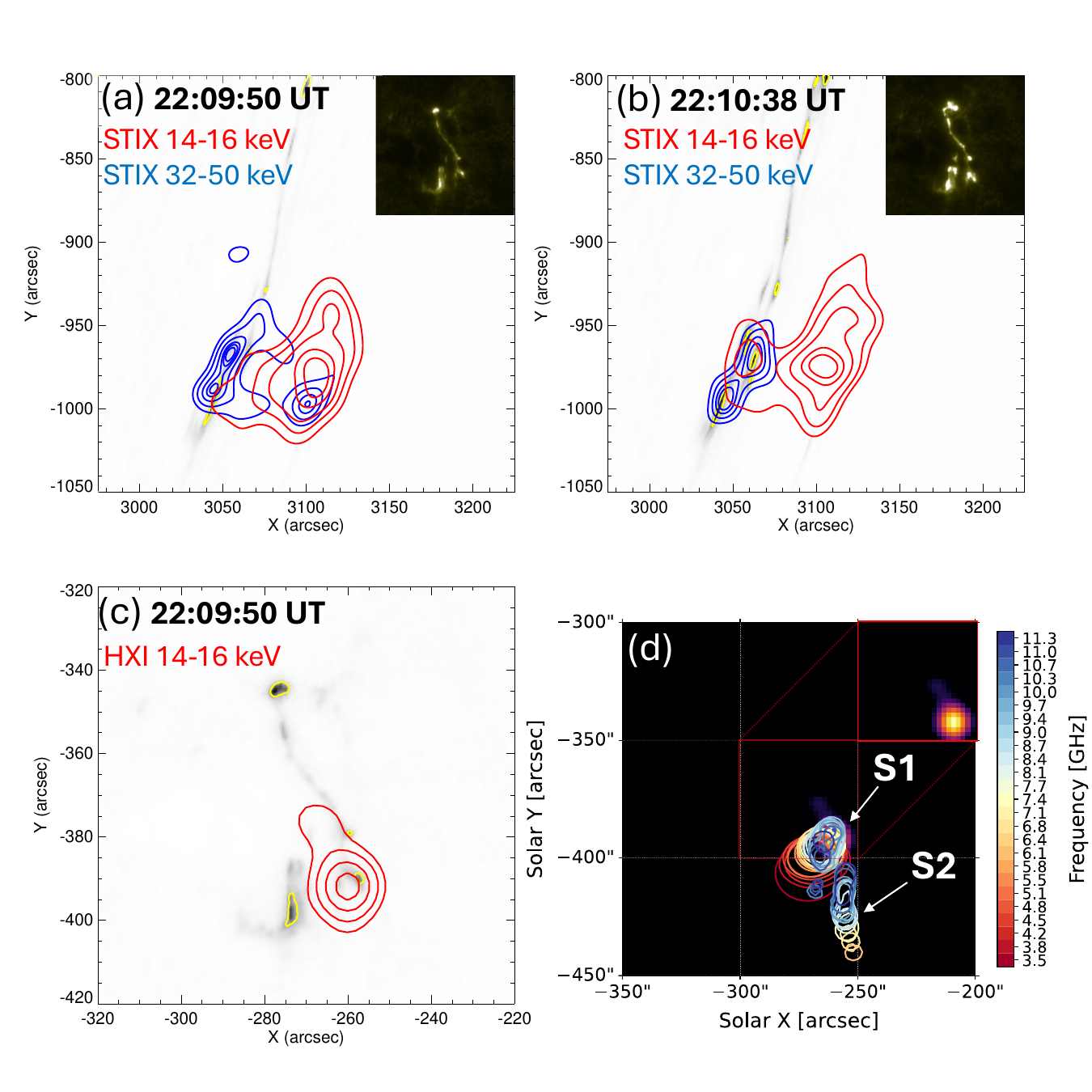}
\caption{(a)–(b) AIA 1600 \AA\ images reprojected to the STIX viewpoint and overlaid with STIX HXR sources. (a) AIA 1600 \AA\ at 22:09:50 UT with 14–16 keV (red) and 32–50 keV (blue) sources integrated over 22:09:04–22:10:04 UT. (b) AIA 1600 \AA\ 22:10:38 UT with the same energy bands integrated over 22:10:30–22:10:50 UT. Blue contours show 15\%, 30\%, 50\%, 70\%, 80\%, and 90\% of the peak intensity in (a), and 15\%, 30\%, 50\%, and 70\% in (b). Red contours show 11\%, 20\%, 35\%, 60\%, and 80\% in both panels. (c) AIA 1600 \AA\ image at 22:09:50 UT with HXI 14–16 keV sources (22:09:04–22:10:04 UT). Red contours show 35\%, 50\%, 70\%, and 90\%. (d) HXI 14–16 keV image from (c) with EOVSA microwave contours at 22:09:54 UT. In (a)–(c), yellow contours show the enhanced regions of the AIA image. Insets in (a) and (b) show the original AIA images, and the inset in (d) shows the HXI image without microwave contours. The related movie for panel (d) starts from 22:07:00 UT to 22:09:54 UT. The background image is fixed to the AIA 131 \AA\ image at 22:08:30 UT, while the image is ASO-S/HXI HXR in panel (d). Note that the microwave contours in panel (d) are shown at 50\% of the peak intensity at each frequency at 22:09:54 UT, whereas those in the animation are fixed at 20\% of the peak intensity at each frequency to show the weaker microwave sources appearing at earlier times.
}\label{fig:fig3}
\end{figure*}

Figures \ref{fig:fig3}(a)–(b) show AIA 1600 \AA\ images reprojected to the STIX viewpoint, overlaid with STIX HXR contours for Phases 0 and 1. Two footpoint sources appear at similar locations in both phases, and the flare-ribbon morphology remains consistent, indicating that the principal reconnection site did not change during this interval. In Phase 0, the non-thermal coronal HXR source is detected, whereas it is not visible in Phase 1. This is consistent with the coronal source peaking earlier and more rapidly than the footpoint emission \citep{Masuda1994,Ishikawa2011}. The coronal source likely persists into Phase 1 but falls below the imaging dynamic range, implying that it is much fainter than the footpoint sources in Phase 1. Figure \ref{fig:fig3}(c) shows the AIA 1600 \AA\ image with HXI HXR contours in Phase 0. Because of low photon counts, the non-thermal 32–50 keV emission could not be reliably imaged. Nevertheless, the thermal looptop location is identified in 14–16 keV. Figure \ref{fig:fig3}(d) overlays the Phase 0 EOVSA microwave contours on the HXR image used in panel (c). EOVSA reveals two non-thermal microwave sources, S1 and S2. S1, which is the main microwave source, is co-spatial with the HXI HXR source and is therefore interpreted as the looptop source. The secondary microwave source, S2, appeared after S1. It became clearly visible by 22:09:54 UT, as shown in the accompanying movie for Figure \ref{fig:fig3}. The nature of S2 is discussed in the next section, and the appearance time of S2 is mentioned in Section \ref{sec:sec4.3}.

\subsection{Secondary Microwave Source and Associated Ejection} \label{sec:sec3.3}
\begin{figure*}[!htbp]
\centering
\includegraphics[width=0.65\textwidth]{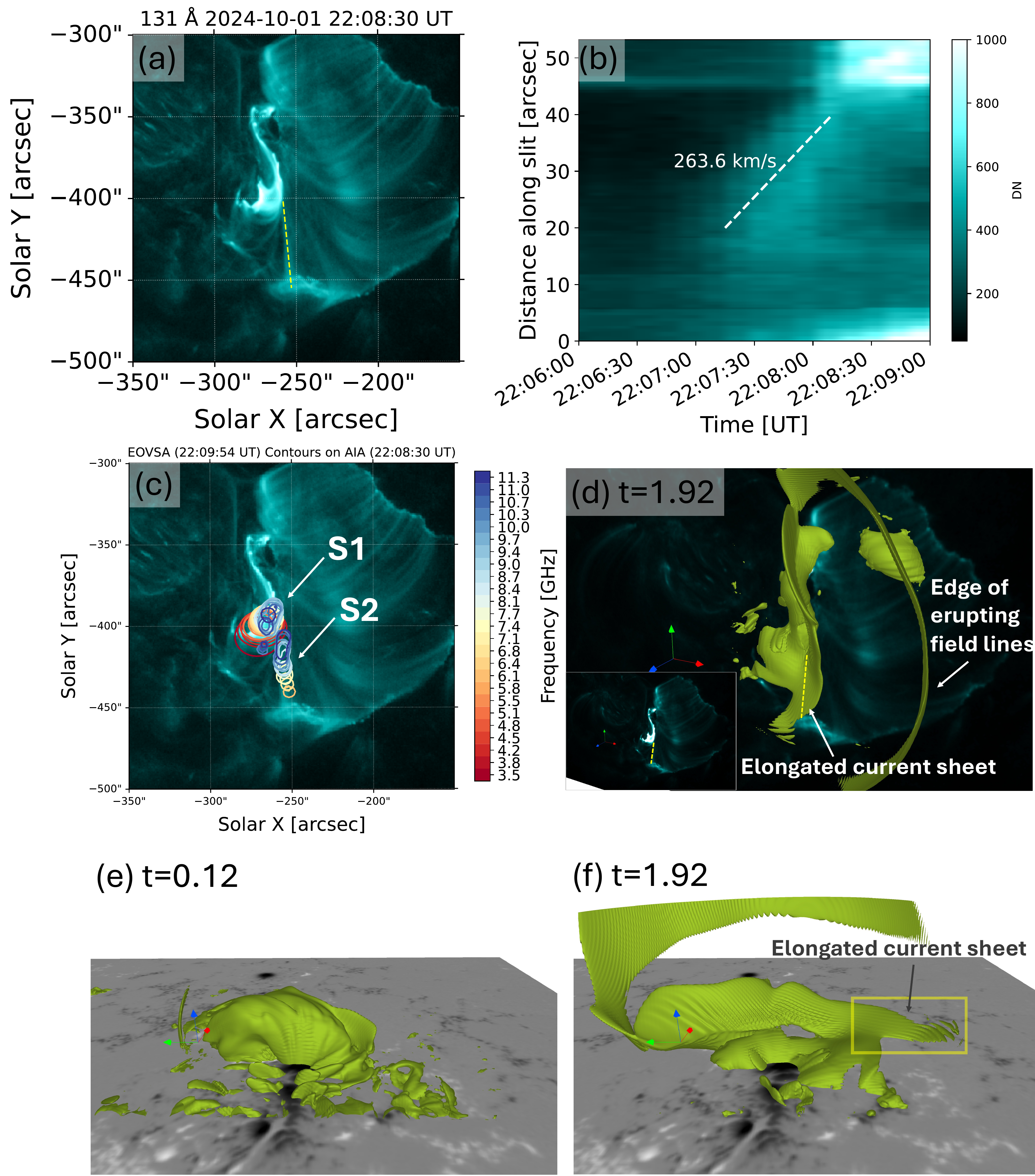}
\caption{(a) AIA 131 \AA\ image at 22:08:30 UT. The yellow line indicates the slit along which the eruption is observed to propagate southward. (b) Time-distance diagram of the AIA 131 \AA\ emission along the slit shown in panel (a), where the distance along the vertical axis is measured from the northern end of the slit at $(x,y)=(-258^{\prime\prime}, -402^{\prime\prime})$. The color scale indicates the intensity of the data number. (c) Same AIA 131 \AA\ image as in panel (a), overlaid with EOVSA microwave contours at 22:09:54 UT. The main microwave source is labeled S1, and the secondary source observed along the eruption is labeled S2. (d) Isosurface of $|\bm{J}|=5$ at $t=1.92$ rendered in the observational viewing geometry during the MHD simulation, overlaid on the AIA 131 \AA\ image at 22:08:30 UT. The bottom-left inset shows the AIA image alone. The yellow dashed line traces the eruption direction. (e)-(f) Isosurface of $|\bm{J}|=5$ at $t=0.12$ and $t=1.92$ in side view. Red, green, and blue arrows in each panel show the $x$, $y$, and $z$ axes in the simulation box. The animation of panel (a) in narrow view from 22:04:30 UT to 22:12:11 UT is available, with EOVSA 7.7 GHz contours at 22:09:54 UT overlaid. The animations of panels (d), (e), and (f) from $t=0.00$ to $2.40$ are available.
}\label{fig:fig4}
\end{figure*}

Figure \ref{fig:fig4}(a) shows the AIA 131 \AA\ image at 22:08:30 UT. The yellow slit marks the path along which a plasma ejection develops, beginning at 22:07:00 UT. 
Figure \ref{fig:fig4}(b) presents the time–distance diagram along this slit. To improve the signal-to-noise ratio, intensities were averaged over a 3-pixel width in the $y$ direction. In this panel, the distance is measured from the northern end of the slit. The ejection track yields a projected speed of 263.6 km s$^{-1}$. 
Figure \ref{fig:fig4}(c) overlays EOVSA microwave contours on the AIA image. The secondary microwave source, S2, lies along the ejection track. This configuration resembles those reported for other flares \citep{Chen2020,Kou2022}, in which the secondary microwave emission arises from the current sheet region at and above the flare arcades. Figure \ref{fig:fig4}(d) shows the 3D isosurface at $|\bm{J}|=5$ at $t=1.92$ overlaid on the AIA 131 \AA\ image at 22:08:30 UT. This surface is used to track the evolution of the current sheet. Two $|\bm{J}|$ structures are evident. One is the edge of the erupting field lines. The other is a concentration around the sigmoidal structure seen in AIA 131 \AA. The yellow dashed line traces the observed eruption path. Figures \ref{fig:fig4}(e) and (f) show that the current sheet elongates southward along the observed eruption path toward the southern region. While the current sheet also evolves elsewhere, the secondary microwave emission is detected only near the southern segment. The physical origin of this localization is discussed in Section \ref{sec:sec4.2}.

\subsection{Estimated Height of S1} \label{sec:sec3.4}
In this section, we estimate the height of S1 above the solar surface by combining the stereoscopic geometry with the STIX and EOVSA imaging. Then we compare the inferred height with the corresponding height of the looptop in the data-constrained MHD simulation.

\begin{figure*}[!htbp]
\centering
\includegraphics[width=0.65\textwidth]{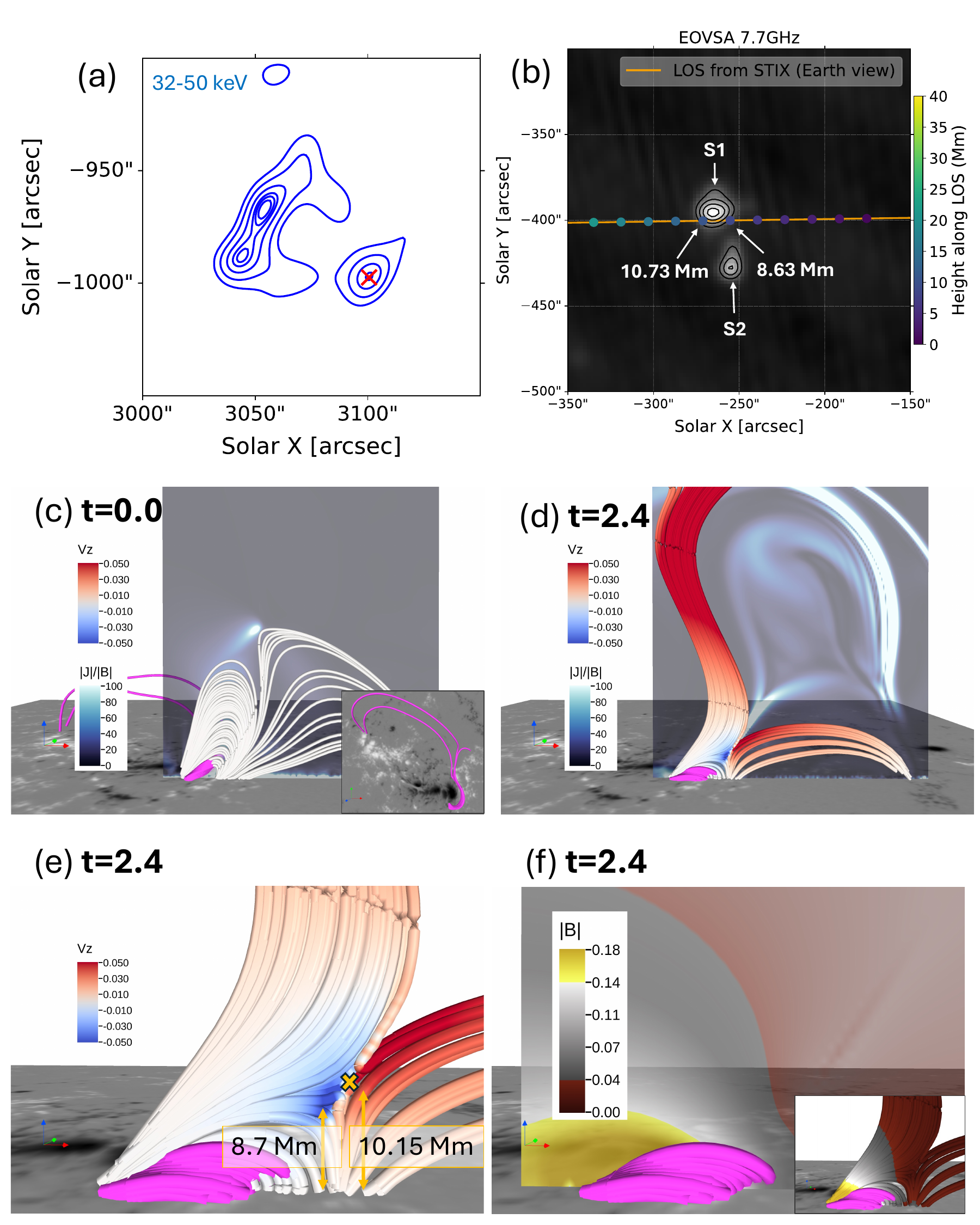}
\caption{(a) STIX 32-50 keV HXR contours (same as Figure \ref{fig:fig3}(a)). Red x marks the brightest point of the coronal source at $(x,y) =(3100^{\prime\prime},-997^{\prime\prime})$. (b) EOVSA 7.7 GHz image at 22:09:54 UT with 30\%, 50\%, 70\%, and 90\% contours. The orange curve is the Earth-view projection of the STIX line of sight through the red x in panel (a), giving the possible projected locations of the STIX coronal source. The color bar shows the corresponding possible height above the solar surface along this line of sight. (c)–(d) Temporal evolution of the field lines (purple) and 2D overlying field lines (colored by $V_z$), associated with observed HXR footpoint sources. Vertical slice shows $|\bm{J}|/|\bm{B}|$. The inset in (c) shows the field lines from above. (e) Zoomed view of (d) without the slice. The orange X marks the reconnection site, and the arrows indicate the heights of the reconnected loops and the magnetic bottle. (f) Same viewpoint and field of view as (e), showing the same field lines. The slice shows $|\bm{B}|$, and the inset shows the 2D overlying field lines colored by $|\bm{B}|$. Red, green, and blue arrows in panels (c)-(f) show the $x$, $y$, and $z$ axes in the simulation box.}
\label{fig:fig5}
\end{figure*}

Figure \ref{fig:fig5}(a) shows the STIX non-thermal HXR image in Phase~0. The red “x” marks the brightest point of the coronal source at $(x,y)=(3100^{\prime\prime},-997^{\prime\prime})$ in the STIX view. Next, we project the line of sight (LoS) of this coronal source into the Earth view. Figure \ref{fig:fig5}(b) displays the EOVSA 7.7 GHz image and contours with the line obtained by projecting the line of sight through the red “x” in the Earth view. This line intersects the S1 microwave source, indicating that the STIX non-thermal HXR source corresponds to the non-thermal looptop source observed by EOVSA. The color scale in panel (b) gives the height above the solar surface along the line. The estimated height of the non-thermal looptop source (S1 microwave and STIX coronal HXR) is 8.63–10.73 Mm. 

Figures \ref{fig:fig5}(c)–(d) show the temporal evolution of the magnetic field. Note that all field lines colored in $V_z$ are projected onto the $y-z$ plane. The purple field lines are traced from the HXR footpoint observed by HXI in Phase 1(see Figure 2(b) in \citet{Matsumoto2026}), and the field lines colored by $V_z$ indicate the 2D overlying field in this region. As discussed in Section \ref{sec:sec3.2}, the main reconnection site responsible for the non-thermal emission is inferred to be the same between the Phases 0 and 1. Therefore, we treat these traced field lines as representative of the magnetic connectivity relevant to the non-thermal emission in Phase 0, and use them to guide our search for the EOVSA microwave sources observed in Phase 0. During the MHD simulation, reconnection forms a current sheet and post-flare loops. Figure \ref{fig:fig5}(e) is a zoom of panel (d). The yellow “x” marks the reconnection site. Two yellow arrows highlight the heights below the reconnection site corresponding to the looptop and the magnetic bottle region, which forms near the lower end of the reconnection current sheet, and both microwave and hard X-ray emissions originate \citep{Chen2024}. In the MHD simulation, the heights relevant to the looptop and the magnetic bottle region range from 8.7 to 10.15 Mm. This range is consistent with the height inferred for the non-thermal looptop source in Figure \ref{fig:fig5}(b). Therefore, both the stereoscopic analysis and the 3D data-constrained MHD simulation provide a realistic 3D magnetic environment and enable source heights to be constrained, even for flares observed away from the limb, where height information is usually limited. Figure \ref{fig:fig5}(f) maps the magnetic field strength $|\bm{B}_{MHD}|=\sqrt{B^2_x+B^2_y+B^2_z}$ from the simulation. Within the magnetic bottle, $|\bm{B}|$ is smaller than 116.6 G, since the normalization uses $B^{*}=2.915\times10^{-1}$~T ($\approx 2915$ G) and the simulated value is $\sim 0.04$ in panel (f). Using the Markov chain Monte Carlo approach (MCMC; see Appendix \ref{AppenA}), we explored the parameter space relevant to the microwave emission. This method samples the posterior probability distribution of the model parameters, thereby allowing us to evaluate not only the most probable parameters but also the associated uncertainties and correlations among the parameters. We infer $|\bm{B}_{MCMC}|=154$ G at the microwave S1 source. Even when the $1 \sigma$ uncertainty range is taken into account ($140.73 \leq |\bm{B}_{MCMC}| \leq 215.52$), the inferred magnitude exceeds the value in the MHD simulation ($|\bm{B}_{MHD}\leq 116.6$ G). For more details of the fitting result, please refer to Appendix \ref{AppenA}.

\section{Discussion} \label{sec:sec4}

\subsection{Different results from MCMC Approach and MHD Simulation}\label{sec:sec4.1}
The result in Section \ref{sec:sec3.4} showed a discrepancy between the magnetic field strength of S1 estimated from the microwave and that obtained from the MHD model. One possible explanation is the limitation of the zero-$\beta$ approximation adopted in the MHD simulation, which uses pseudo density and neglects plasma pressure. Because compression near the reconnection outflow and looptop are not fully captured in our model, the local magnetic field strength in the actual source region could be underestimated, possibly contributing to the discrepancy. Another possible factor is that the HMI magnetograms used as the bottom boundary in the MHD simulation may not fully resolve strong-field structures at small scales. Comparisons with higher-resolution observations have reported that HMI field strengths are lower in active regions \citep{Jin2023,Beck2025}. This implies that HMI may underestimate the photospheric magnetic field strength in unresolved small-scale features. However, because the large-scale magnetic structure remains broadly consistent across these observations, such differences do not necessarily bring a comparable increase in the coronal magnetic field strength. We also checked this possibility by performing additional MHD simulations with magnetograms degraded to different spatial resolutions. The results showed no clear trend, indicating that a higher numerical resolution would not produce magnetic fields around the looptop and current sheet comparable to the EOVSA-derived value. Thus, the discrepancy is unlikely to be explained by the spatial resolution of the MHD simulation.

\begin{figure}[!htbp]
\plotone{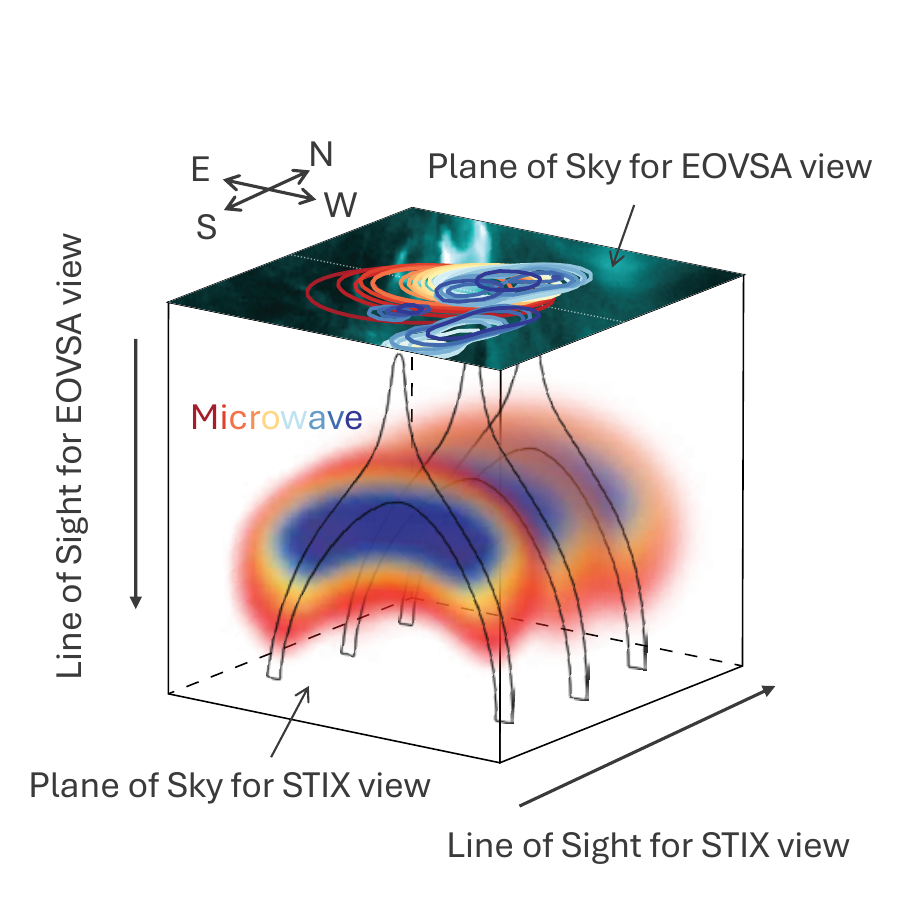}
\caption{Schematic illustration of the viewing geometry used to explain why the magnetic field strength inferred from the microwave emission in the X7.1 flare exceeds the field in the MHD model. The top surface shows the plane for the EOVSA view, while the front face shows the same structure as viewed from the STIX view.
}\label{fig:fig6}
\end{figure}

A plausible and likely important explanation is an effect of inhomogeneous magnetic field distribution along the LoS for the microwave emission (Figure \ref{fig:fig6}), as pointed out earlier \citet{Kaltman2026}. In view of the geometry for the X7.1 flare (top surface of Figure \ref{fig:fig6}), the observed plane of sky corresponds to looking down onto the flare arcade. The LoS roughly follows the loops and integrates the microwave emissivity over a broad height range. This yields a LoS integrated value weighted toward a stronger magnetic field. Indeed, Figure 5 of \citet{Chen2024} shows that the microwave emission is not confined to a compact looptop source, but is distributed over an extended region spanning from the loop legs to the looptop. By contrast, in a limb-view geometry (front face of Figure \ref{fig:fig6}), we see the loops from the side, so the LoS cuts them transversely at nearly one height, sampling a much narrower height range and producing an estimate closer to the local magnetic field. This LoS geometry effect naturally explains why $|\bm{B}|$ estimated from the microwave for the X7.1 flare exceeds the MHD field, whereas the flare at the limb tends to show better agreement \citep{Chen2020}.

\subsection{Origin and Estimated Height of S2} \label{sec:sec4.2}
\begin{figure*}[!htbp]
\centering
\includegraphics[width=0.8\textwidth]{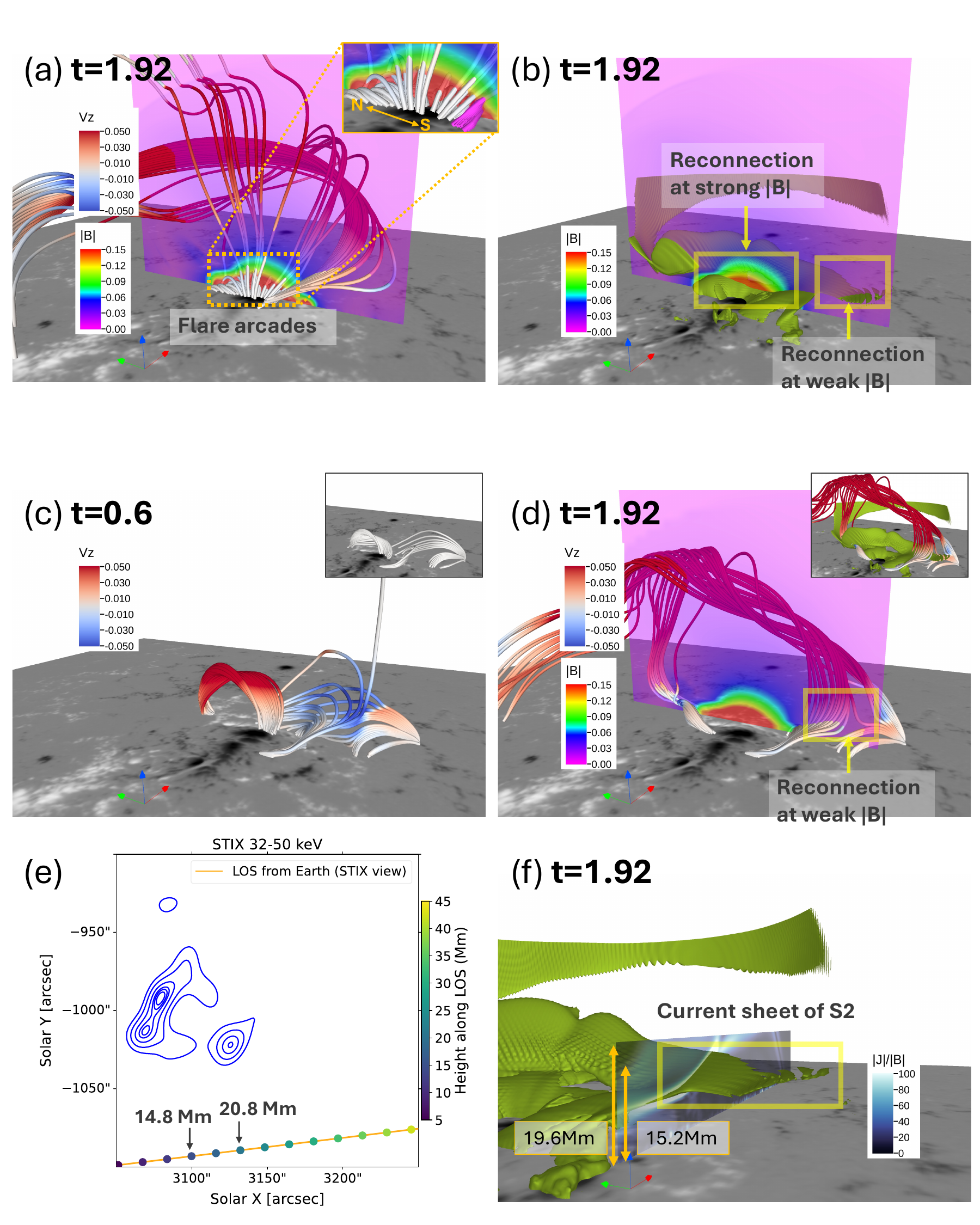}
\caption{(a) Field lines colored in $V_z$ and cross-section of $|\bm{B}|$ at $t=1.92$. Small inset shows the strong $|\bm{B}|$ region, corresponding to the flare arcades. The purple field lines shown in the inset are traced from the same location as the purple field lines in Figure \ref{fig:fig5}. (b) 3D isosurface at $|\bm{J}|=5$ and cross-section of $|\bm{B}|$ at $t=1.92$ (c)-(d) Temporal evolution of field lines associated with the current sheet of S2. Small insets in (c) and (d) show the field lines at $t=0.0$ (NLFFF) and field lines with isosurface at $|\bm{J}|=5$ at $t=1.92$, respectively. (e) STIX HXR 32-50 keV contours same as in Figure \ref{fig:fig3} (a). The orange line shows the LoS from the EOVSA viewpoint through the S2 source, projected into the STIX view. The coordinates of S2 are $(x,y) =(-254^{\prime\prime},-428^{\prime\prime})$ in Figure \ref{fig:fig4}(c). (f) 3D isosurface at $|\bm{J}|=5$ and cross-section of $|\bm{J}|/|\bm{B}|$ to focus on S2. The arrows indicate the height used for comparison with panel (e). Red, green, and blue arrows in panels (a)-(d) and (f) show the $x$, $y$, and $z$ axes in the simulation box.
}\label{fig:fig7}
\end{figure*}

As described in Section \ref{sec:sec3.3}, the data-constrained MHD simulation shows that the current sheet grows spatially not only toward the S2 direction but also over a broader volume. Nevertheless, the secondary microwave emission is detected only near the southern current sheet. To understand this localization, Figures \ref{fig:fig7}(a) and (b) show the field lines, the vertical distribution of $|\bm{B}|$, and the $|\bm{J}|=5$ isosurface at $t=1.92$. The small inset in Figure \ref{fig:fig7}(a) shows that the reconnected field lines of S1, colored in purple, are part of the flare arcades. These field lines lie on the southern side of the flare arcades, which extend in the north-south direction. This is consistent with the thermal looptop emission extended along the north-south direction in Figure \ref{fig:fig3}(c). It indicates that the flare arcades host the thermal looptop source while the field lines most relevant to the non-thermal emission are concentrated in the south. In Figure \ref{fig:fig7}(a), the strong magnetic field region corresponds to the flare arcades, where the non-thermal looptop source, S1, was observed. The reason why a secondary microwave component is not clearly detected from the vertically extended current sheet above the flare arcades in Figure \ref{fig:fig7}(b) is that the secondary source is superposed on the much brighter looptop emission from the flare arcade. As a result, the weak secondary source is difficult to detect and separate.

We next consider why the secondary microwave emission is detected specifically near the current sheet associated with S2. Figures \ref{fig:fig7}(c) and (d) show the temporal evolution of the field lines relevant to the S2. As the northern twisted field erupts, it reconnects with the surrounding field lines. In particular, field lines rooted in the southern negative polarity region are converted into twisted field lines through the reconnection, and their footpoints coincide with the AIA 1600 \AA\ brightening region, as seen in Figure \ref{fig:fig2}(d) and reported by \citet{Matsumoto2025b}. Moreover, Figure \ref{fig:fig7}(d) shows that the reconnection associated with S2 takes place in a much weaker magnetic field region than that associated with S1. From these results, we argue that reconnection in this event may occur under different magnetic field strengths, yielding a more complex picture than the single system by \citet{Kou2022}, who interpreted both the main and secondary microwave sources as arising within a single reconnection system.

We project the LoS of S2 into the STIX view to estimate its height. Figure \ref{fig:fig7}(e) overlays the STIX HXR image with the LoS to S2 as seen from the EOVSA view, projected into the STIX view plane. This projected line gives the possible heights and locations of S2. The line does not intersect the STIX coronal HXR source. This is consistent with S2 being invisible in the STIX HXR image. Assuming that S2 lies south of S1 in the image because S2 is spatially resolved south of S1 in the EOVSA image, the inferred height is $\sim$14.8-20.8 Mm. Figure \ref{fig:fig7}(f) shows the isosurface at $|\bm{J}|=5$ and cross-section of $|\bm{J}|/|\bm{B}|$ to locate the corresponding height range in the simulated current sheet. The height range $\sim$15.2--19.6 Mm indicated in Figure \ref{fig:fig7}(f) lies within the current sheet associated with S2 and above the reconnected loops in Figure \ref{fig:fig5}(e). Therefore, the stereoscopic analysis and the data-constrained MHD simulation suggest that S2 is located southward of the main reconnection site associated with S1 and originates from a higher altitude than S1.

\subsection{Possible Radiating Conditions for S2}\label{sec:sec4.3}

We further investigate what physical conditions are required to produce S2 through forward modeling. Because S2 has a relatively low signal-to-noise ratio and its spectrum appears to contain multiple components (Figure \ref{fig:apdixS2}), it should be emphasized that a unique and reliable fit with a simple homogeneous source model is not possible. Nevertheless, the brightness temperature of S2 exceeds 10 MK (Appendix \ref{AppenB}), strongly suggesting that S2 is the non-thermal source. We do not attempt to derive a definitive parameter set for S2. Instead, we present only a representative model. The solutions with a fixed low-energy cutoff of $E_{\min} = 10$ keV tend to require an unusually high non-thermal electron density and an implausibly low thermal density compared with S1 (Case 1 in Table \ref{tab:fitting}). This solution is less consistent with the absence of a clear STIX HXR counterpart at the location of S2 (Section \ref{sec:sec4.2}) and with the differential emission measure analysis (Appendix \ref{AppenB}).

 By contrast, a representative solution with a higher $E_{\min}$ can reproduce the observed S2 spectrum without such extreme thermal densities (Case 2 in Table \ref{tab:fitting}). In this sense, the S2 emission may be more naturally explained by a non-thermal electron population weighted toward higher energies than in S1. This scenario is supported by some observations. Observationally, plasma ejection is seen from the region of  S1 toward that of S2 in the southern direction (Section \ref{sec:sec3.3}), meaning the spatial relationship. Moreover, the appearance of S2 was during the observed plasma ejection, as shown in Appendix \ref{AppenC}, meaning the temporal relationship. This representative solution could reflect transport of pre-accelerated electrons from the main source region, additional acceleration along the southern current sheet, or a combination of both. Note that the present results do not allow these possibilities to be distinguished, and the inferred parameters should not be regarded as unique. A more exact nature of the secondary microwave source will require microwave imaging spectroscopy with substantially higher image dynamic range and fidelity, such as Frequency Agile Solar Radiotelescope (FASR).

\subsection{Acceleration Condition for S1 and S2}\label{sec:sec4.4}

The efficiency of electron acceleration in magnetic reconnection can depend on magnetization $\sigma$, which is the ratio of magnetic energy density to enthalpy density, the plasma beta, and the relative strength of the guide field ($B_g$) to the reconnecting field ($B_r$) \citep{Li2015,Guo2016,Dahlin2016,Li2017,Ball2018}. Since the MHD simulation used in this paper is based on the zero-$\beta$ approximation, we cannot evaluate the plasma beta in the reconnection regions. On the other hand, the 3D magnetic field obtained from the simulation allows us to examine the relative strength of the guide field. In this section, we evaluate the ratio of the guide field to the reconnecting field in the reconnection regions associated with S1 and S2.

For the main reconnection region associated with S1, the current sheet structure related to the reconnection is distributed along the PIL, which is roughly parallel to the $y$ direction in the simulation box (e.g., Figure \ref{fig:fig4}(d)). In the $x-z$ plane, the simulation clearly shows reconnection geometry similar to the CSHKP model \citep{Carmichael1964,Sturrock1966,Hirayama1974,Kopp1976}, as shown in Figure 2 of \citet{Matsumoto2025b}. Therefore, taking the $x-z$ plane as the reconnection plane, we approximate $B_r = \sqrt{B_x^2 + B_z^2}$ and $B_g = |B_y|$. We apply the same approximation to S2. As shown in Section \ref{sec:sec3.3}, S2 is associated with the southern current sheet structure that appears along the southward plasma ejection. This structure is also distributed roughly along the $y$ direction, in the same sense as the current sheet structure in S1. Thus, for S2, we also regard $B_r$ and $B_g$ as $\sqrt{B_x^2 + B_z^2}$ and $B_g = |B_y|$, respectively. Note that this definition does not determine the exact local coordinate system for the guide and reconnecting magnetic field lines. Instead, it provides an approximate diagnosis for comparing the reconnection environments of S1 and S2 on a common basis.

\begin{figure*}[!htbp]
\centering
\includegraphics[width=0.6\textwidth]{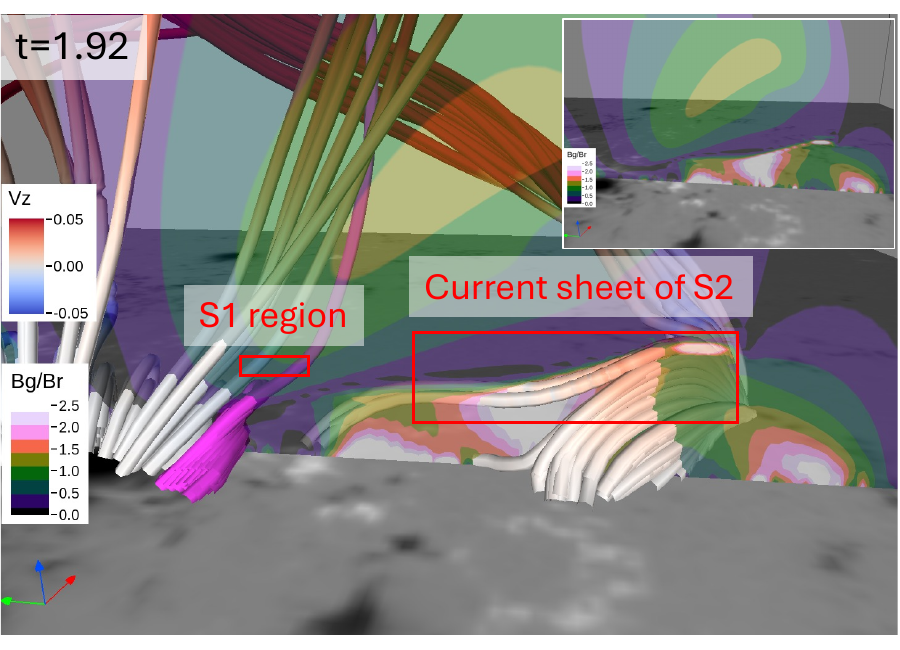}
\caption{Magnetic field lines colored in $V_z$ and cross-section of $B_g/B_r$ at $t=1.92$. Purple field lines show the magnetic field lines of S1, which are traced from the same region as Figure \ref{fig:fig5}. For S1 region, the red square shows the height of 8.7 and 10.15 Mm, which are indicated in Figure \ref{fig:fig5}(e). The small inset shows only the cross-section of $B_g/B_r$. Red, green, and blue arrows in each panel show the $x$, $y$, and $z$ axes in the simulation box.}\label{fig:fig8}
\end{figure*}

Figure \ref{fig:fig8} shows the spatial distribution of $B_g/B_r$ around the reconnection regions. Around S1, $B_g/B_r$ is finite, with typical values of approximately $0.5$--$1$. This suggests that the reconnection region associated with S1 may provide conditions favorable for efficient electron acceleration \citep{Dahlin2016}. This interpretation is consistent with the observations, in which the non-thermal HXR and microwave sources are observed. In contrast, the southern current sheet region associated with S2 shows larger values of $B_g/B_r$ than the S1 region, with values generally exceeding $1$ and locally exceeding $2.5$. This result suggests that, around S2, the guide field is relatively stronger than the reconnecting field, and that the local reconnection condition may be less favorable for efficient electron acceleration than in the S1 region \citep{Dahlin2016}. Therefore, the higher $E_{\min}$ inferred for S2 may not necessarily indicate efficient local acceleration at S2. Instead, it may reflect the transport of higher energy electrons accelerated around S1 along the southward plasma ejection. However, the present analysis alone cannot quantitatively distinguish whether the electron distribution at S2 is produced mainly by transport from S1 or includes additional acceleration in the southern current sheet region. Addressing this issue requires test particle simulations based on the MHD fields or modeling that solves the transport equation for electrons. We leave such quantitative analysis for future work.
\section{Summary} \label{sec:sec5}
We investigated the X7.1 flare in NOAA AR 13842 by combining stereoscopic HXR observations from Solar Orbiter/STIX and ASO-S/HXI with the microwave imaging from EOVSA, and by comparing the results with a 3D data-constrained MHD simulation that uses an NLFFF extrapolation as the initial condition. Unlike previous studies that reconstruct the 3D X-ray source geometry entirely from the observed morphology on the projected sky plane \citep{Ryan2024a,Ryan2024b}, this research combines stereoscopic non-thermal HXR and microwave imagings, and a data-constrained MHD simulation to further constrain the source heights and investigate the detailed dynamics of non-thermal emissions.

For efficient comparison of the simulation with the observations, we set an appropriate simulation time interval based on the distance over which magnetic connectivity changes and the distance to the remote brightening in AIA 1600 \AA. This constraint supports the time selection adopted in this paper. In the observations, the primary non-thermal microwave source (S1) is located at the same position as the HXI looptop HXR source and is also consistent with the STIX non-thermal coronal HXR source under the stereoscopic analysis, indicating that these emissions originate from the same coronal structure. The height of S1 obtained from the stereoscopic observation agrees with that of the looptop and magnetic bottle in the MHD simulation. However, the magnetic field strength inferred from the EOVSA spectra exceeds the local value of the MHD simulation. Possible explanations include the zero-$\beta$ approximation adopted in the simulation, unresolved strong-field structure in the photospheric boundary data, and the different emissions from inhomogeneous magnetic field strengths are mixed up along the LoS. 

EOVSA captured the impulsive phase and revealed a secondary non-thermal microwave source (S2) aligned with a southward plasma ejection seen in AIA 131 \AA. In the simulation, the current sheet elongates along the observed ejection toward the S2 region. Moreover, the simulation suggests that reconnection proceeds at multiple locations with different magnetic field strengths during the eruption. In this picture, the looptop source S1 is associated with reconnection in a stronger magnetic field region near the flare arcade, whereas S2 is linked to reconnection in a weaker magnetic field region. Furthermore, the stereoscopic analysis and the data-constrained MHD simulation suggest that S2 lies in the south of the main reconnection site related to S1 and originates at a higher altitude than that of S1. 

We then investigated the physical conditions required for S2 to produce the observed microwave emission. Overall, the physical conditions of S2 are more plausibly explained by a solution with a higher $E_{\min}$ than by one requiring an extremely dense non-thermal electron population at S2 than that at S1. Considering the observational evidence of plasma ejection from S1 toward S2, it is plausible that high-energy electrons accelerated near S1, or re-accelerated on the S2 side, were present along the southern current sheet. Nevertheless, this interpretation is not unique at present. Additional constraints from HXR spectral analysis and 3D forward modeling and next-generation radio observations, such as FASR, will be required to further reduce the remaining parameter degeneracy. We also found that $B_g/B_r$ is approximately $0.5$--$1$ around S1 but generally exceeds $1$ along the southern current sheet associated with S2. This suggests that S1 provides more favorable conditions for efficient electron acceleration, whereas the higher $E_{\min}$ inferred for S2 may reflect transport of higher energy electrons from S1 rather than efficient local acceleration.

Compared with stereoscopic observations, the data-constrained MHD simulation can provide a realistic 3D magnetic field evolution to interpret the non-thermal emission. Future work will extend this approach to finite-$\beta$ data-constrained MHD simulations in a spherical coordinate system, which are needed to capture coronal structure and curvature in the large scale when tracking flares over long time scales, and will couple them with macroscopic particle simulations and HXR spectral analysis to place more realistic constraints on particle acceleration.

\begin{acknowledgments}
this paper is supported by NASA grants 80NSSC23K0406, 80NSSC21K1671, 80NSSC21K0003, 80NSCC24M0174, 80NSSC24K1242, and NSF grants AST-2204384,  AGS-2145253, 2149748, 2206424, 2309939 and 2401229. The 3D visualizations were produced using VAPOR (\href{http://www.vapor.ucar.edu}{\texttt{www.vapor.ucar.edu}}), a product of the National Center for Atmospheric Research \citep{Li2019}. All calculations of the MHD simulations in this paper were performed using the computing facilities of the High Performance Computing Center (HPCC) at the New Jersey Institute of Technology. The authors acknowledge the usage of publicly open data from the ASO-S mission, which is supported by the Strategic Priority Research Program on Space Science, the Chinese Academy of Sciences, Grant No. XDA15320000. Solar Orbiter is a mission of international cooperation between ESA and NASA, operated by ESA \citep{Muller2020}. The STIX instrument is an international collaboration between Switzerland, Poland, France, Czech Republic, Germany, Austria, Ireland, and Italy. SK is supported by Swiss PRODEX grant for STIX. The HXI and STIX datasets analyzed in this paper are publicly available from the Data Archives (\url{http://aso-s.pmo.ac.cn/sodc/dataArchive.jsp} and \url{https://soar.esac.esa.int/soar/}).
\end{acknowledgments}

\appendix
Appendix \ref{AppenA} provides the fitting details for S1, and Appendix \ref{AppenB} presents those for S2. We first summarize the results of the microwave forward fitting for S1 and S2. 
\begin{table*}[!htbp]
\centering
\caption{Summary of the fitting parameters for S1 (looptop source) and S2 (secondary source).}
\label{tab:fitting}
\begin{tabular}{lccccccccc}
\toprule
Source 
& $|B|$ [G] 
& $\log_{10}(n_{\mathrm{ele}})$ [cm$^{-3}$]
& $\delta$ 
& $E_{\min}$ [keV] 
& $E_{\max}$ [MeV] 
& $\theta$ [deg] 
& $\log_{10}(n_{\mathrm{th}})$ [cm$^{-3}$] 
& depth [arcsec] \\
\midrule
S1 
& 154.9$^{a}$ & 9.7$^{a}$ & 4.2$^{a}$ & 10.0$^{b}$ & 10.0$^{b}$ & 50.1$^{a}$ & 10.9$^{a}$ & 15.0$^{b}$ \\
S2 (Case 1)
& 100.0$^{b}$ & 10.9$^{a}$ & 4.5$^{a}$ & 10.0$^{b}$ & 10.0$^{b}$ & 85.8$^{a}$ & 7.1$^{a}$ & 15.0$^{b}$ \\
S2 (Case 2)
& 100.0$^{b}$ & 8.9$^{a}$ & 4.5$^{a}$ & 37.0$^{a}$ & 10.0$^{b}$ & 87.8$^{a}$ & 10.9$^{a}$ & 15.0$^{b}$ \\
\bottomrule
\end{tabular}

\vspace{2mm}
\begin{minipage}{0.95\textwidth}
\small
\textit{Notes.} Superscript $^{a}$ indicates a free parameter, while superscript $^{b}$ indicates a fixed parameter during the forward fitting. Magnetic field strength $|B|$, non-thermal electron number density $n_{\mathrm{ele}}$, spectral index $\delta$, low and high energy cutoffs for non-thermal electrons ($E_{\min}$ and $E_{\max}$), viewing angle $\theta$ with respect to the direction of the magnetic fields, thermal plasma number density $n_{\mathrm{th}}$, and source depth along the line of sight. Note that the reported parameters of S1 are corresponding to the highest posterior sample in the MCMC chain, rather than the peak or median of each marginalized one-dimensional distribution in Figure \ref{fig:apdix}.
\end{minipage}
\end{table*}
\section{Spectral Fitting for S1}
\label{AppenA}
\begin{figure*}[!htbp]
\centering
\includegraphics[width=0.7\textwidth]{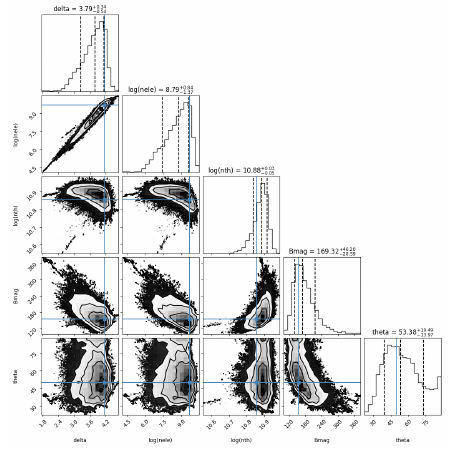}
\caption{MCMC analysis for the spectral fit of the EOVSA data at S1, $(x,y)=(-266^{\prime\prime},-396^{\prime\prime})$, as mentioned in Section \ref{sec:sec3.4}. The diagonal panels show the distributions of each fitted parameter, and the other panels show how pairs of parameters are correlated. The dashed lines in the histograms indicate the 16 percent, 50 percent (median), and 84 percent quantiles corresponding to 1 $\sigma$ interval from the median. The values shown above each histogram indicate the median, with superscripts and subscripts indicating offsets from the median to the 84 and 16 percentiles, respectively. Blue point shows the location of fitting result (highest posterior sample in MCMC), where $\delta$, $log (n_{\rm ele})$, $log (n_{\rm th})$, $B_{\rm mag}$, $\theta$ are 4.16, 9.69, 10.86, 154.89, and 50.10, respectively. The fitting result is summarized in Table \ref{tab:fitting}.
}\label{fig:apdix}
\end{figure*}

Using MCMC analysis, we can investigate the parameter space for microwave spectrum, including the spectral index (delta: $\delta$), the density of non-thermal electrons ($log (n_{\rm ele})$), the density of thermal plasma ($log (n_{\rm th})$), the strength of magnetic fields ($B_{\rm mag}=|B|$), and view angle $\theta$ with respect to the direction of the magnetic fields. More detailed information on the method is written in \citet{Chen2020,Chen2021}. We set the column depth to $15^{\prime\prime}$ ($\approx 10.8$~Mm) based on the looptop height inferred from the MHD simulation in Section \ref{sec:sec3.4}. Figure \ref{fig:apdix} shows the corner plot of the parameter distributions obtained from the MCMC analysis. The diagonal panels present the one-dimensional distributions of each parameter, while the off-diagonal panels show the corresponding two-dimensional distributions. This allows us to evaluate not only how well each parameter is constrained, but also how the parameters are correlated or degenerate with one another.
\section{Spectral Fitting for S2}
\label{AppenB}
\begin{figure*}[!htbp]
\centering
\includegraphics[width=0.95\textwidth]{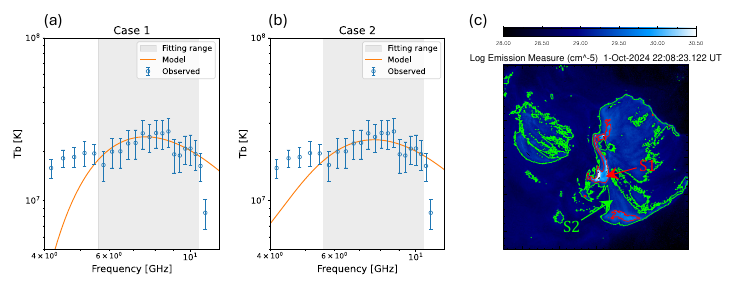}
\caption{(a)-(b) The comparison in observation and modeled spectrum for S2, based on the fitting parameters, as summarized in Table \ref{tab:fitting}. The fitting frequency range is carefully chosen to focus on the observed S2 frequency. Case 1 gives a fit with fixed $E_{\min}$, whereas Case 2 is obtained by freeing $E_{\min}$. Both cases reproduce the observed spectrum. (c) Emission measure map at 22:08:23 UT, when the plasma ejection was observed (Section \ref{sec:sec3.3}). The map shows the emission measure on a logarithmic scale ($\log_{10}(EM)$). Red and green contours indicate the emission measure in the S1 and S2 regions, where the contours are $10^{29}$ and $10^{29.5} \mathrm{cm^{-5}}$, respectively. 
}\label{fig:apdixS2}
\end{figure*}

We first attempted to constrain the model parameters using MCMC, as was done for S1. However, the solutions became highly multimodal with parameter degeneracies, making it difficult to obtain a stable physical interpretation from the global posterior distribution. Therefore, for S2, we used the Nelder-Mead and basinhopping algorithms, accessed through the \texttt{pygsfit} \footnote{\url{https://github.com/suncasa/pygsfit/}} package, to search for the model parameters that reproduce the observed spectrum. The Nelder-Mead method performs a local optimization around the current parameter set, whereas basinhopping repeatedly perturbs the parameters and re-applies the Nelder-Mead minimization in a different region of parameter space. This strategy reduces the risk of being trapped in a local minimum and increases the likelihood of finding a better overall solution than Nelder-Mead alone. For the spectral fitting, the center coordinates $(x_{\mathrm{cen}}, y_{\mathrm{cen}})=(-246^{\prime\prime}, -433^{\prime\prime})$ and spatial widths of $x_{\mathrm{wid}}=36^{\prime\prime}$ and $y_{\mathrm{wid}}=54^{\prime\prime}$ are selected to cover the S2 source. The magnetic field strength is fixed at 100 G, based on the value at the extended current sheet of S2. We note that even when a lower magnetic field strength (e.g., $60$ G) was assumed, the discussion in Section \ref{sec:sec4.3} remained unchanged. The fitting frequency range is selected to focus on the frequencies where S2 is clearly observed, as shown in gray region of Figure \ref{fig:apdixS2}. For S2, the source depth along the line of sight was estimated approximately from the vertical thickness of the extended current sheet, defined by the isosurface of $|\bm{J}|=5$. We note that this thickness depends on the adopted threshold value of $|\bm{J}|$. However, the inferred spectral parameters are expected to be only weakly sensitive to the optical depth \citep{Gary2013, Kuroda2020}. In practice, we tested several plausible values and found that the fitting results changed only marginally.

Figures \ref{fig:apdixS2} (a) and (b) show the comparison between the observed and modeled spectra for S2 using the fitting parameters summarized in Table \ref{tab:fitting}. The gyrosynchrotron
radiation calculation for panels (a) and (b) is conducted based on the fast gyrosynchrotron codes \citep{Fleishman2010}. Case 1 gives a solution with a non-thermal electron density with fixed $E_{\min}$, while Case 2 is obtained by allowing \(E_{\min}\) to vary as a free parameter. Although these two cases yield different parameter sets, both reproduce the observed spectrum over the S2 frequency range, suggesting a degeneracy among the fitting parameters. To investigate whether $10^{3}$ higher thermal plasma density in S1 than that in S2 is realistic (Case 1), a differential emission measure was performed using AIA observations (94, 131, 171, 193, 211, and 335 \AA), based on \citet{Cheung2015}. We calculated the emission measure (EM), which is defined as the LoS integral of the squared thermal electron density ($n_e$), $EM = \int n_{e}^2 \, ds$, and is expressed in units of $\mathrm{cm^{-5}}$. The code implementing this method is open source and publicly available.\footnote{\url{https://www.lmsal.com/~cheung/AIA/tutorial_dem/}} Figure \ref{fig:apdixS2} (c) indicates that the emission measure in the S1 region is only about $10^{0.5}$ times larger than that in the S2 region. This suggests that it is highly unlikely that the number of thermal plasma in S1 is as much as $10^{3}$ times greater than in S2. On the other hand, the solution with $E_{\min} \sim 37$ keV (Case 2) required a lower non-thermal electron density than S1 and comparable thermal plasma density to S1.

\section{Appearance of S2}
\label{AppenC}

\begin{figure*}[!htbp]
\centering
\includegraphics[width=0.9\textwidth]{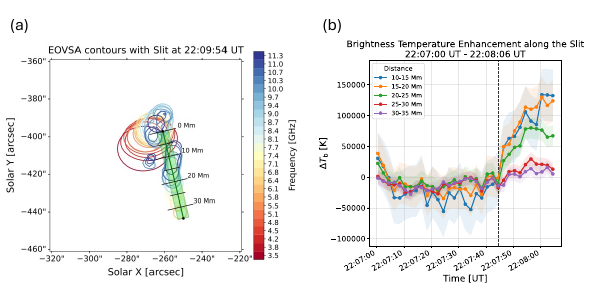}
\caption{(a) Microwave observations observed by EOVSA at 22:09:54 UT, identical to those shown in Figures \ref{fig:fig3} (d) and \ref{fig:fig4}(c), with the slit placed along S2. The slit width is 3 pixels, corresponding to about 6 arcsec (4.35 Mm). The short tick marks along the slit indicate 5 Mm intervals.
(b) Temporal evolution of the background-subtracted brightness temperature enhancement ($\Delta T_b$) along the slit. The brightness temperature, which is $T$ (K), was averaged over the frequency bands and over each 5 Mm segment along the slit. Then, for each distance bin, the mean brightness temperature during the background interval (22:06:00 - 22:07:00 UT) was subtracted. The 0-10 Mm range is excluded because it partially overlaps the bright S1 source. The vertical dashed line marks 22:07:44 UT. The lightly shaded region around each curve indicates the ($\pm 1\sigma$) standard deviation measured during the background interval for the corresponding distance bin.
}\label{fig:apdixEOVSA}
\end{figure*}

In Section \ref{sec:sec3.3}, S2 microwave source is spatially associated with plasma ejection observed by AIA 131 \AA. In this appendix, we further examine their temporal relationship, as mentioned in Section \ref{sec:sec4.3}. Figure \ref{fig:apdixEOVSA}(a) shows the same EOVSA contours as those in Figures \ref{fig:fig3} (d) and \ref{fig:fig4}(c), together with a slit placed along S2. Figure \ref{fig:apdixEOVSA}(b) shows the temporal evolution of the brightness temperature enhancement ($\Delta T_b$) along this slit. At each time step, the brightness temperature was averaged over the frequency bands and then spatially averaged in 5 Mm intervals along the slit. Next, the mean brightness temperature during the background interval of 22:06:00-22:07:00 UT was subtracted for each distance bin to calculate $\Delta T_b$. As shown in Figure \ref{fig:apdixEOVSA}(b), $\Delta T_b$ started to increase at around 22:07:44 UT, as indicated by the dashed line. This timing coincides with the plasma ejection shown in Figure \ref{fig:fig4}(b), suggesting that the appearance of S2 was temporally associated with this plasma ejection.

\bibliography{sample631}{}
\bibliographystyle{aasjournal}
%




\end{document}